\documentclass[twocolumn]{aastex62}

\usepackage{amsmath}
\usepackage{mathtools}
\usepackage[T1]{fontenc}
\usepackage{apjfonts}

\newcommand{\code}[1]{\texttt{#1}}
\newcommand{\mesa}{\code{MESA}}
\newcommand{\MESA}{\mesa}
\newcommand{\athena}{\code{Athena++}}
\DeclarePairedDelimiter{\paren}{\lparen}{\rparen}

\newlength{\apjcolwidth}
\newlength{\figwidth}
\newlength{\doublewide}
\begin{document}
\title{
Remnants of Subdwarf Helium Donor Stars Ejected from Close Binaries with Thermonuclear Supernovae
}

\author[0000-0002-4791-6724]{Evan B. Bauer}
\affiliation{Kavli Institute for Theoretical Physics, University of California, Santa Barbara, CA 93106, USA}
\correspondingauthor{Evan B. Bauer}
\email{ebauer@kitp.ucsb.edu}

\author{Christopher J. White} 
\affiliation{Kavli Institute for Theoretical Physics, University of California, Santa Barbara, CA 93106, USA}

\author{Lars Bildsten}
\affiliation{Kavli Institute for Theoretical Physics, University of California, Santa Barbara, CA 93106, USA}
\affiliation{Department of Physics, University of California, Santa Barbara, CA 93106, USA}

\begin{abstract}
Some binary systems composed of a white dwarf (WD) and a hot subdwarf
(sdB) helium star will make contact within the helium
burning lifetime of the sdB star. 
The accreted helium on the  WD inevitably undergoes a thermonuclear
instability, causing a  detonation that is expected to transition into
the WD core and lead to a thermonuclear
supernova while the donor orbits nearby with high velocity.
Motivated by the recent discovery of fast-moving objects that occupy
unusual locations on the HR diagram, we explore
the impact of the thermonuclear supernovae on the donors in 
this specific double detonation scenario. 
We use \mesa\ to model the binary up to the moment of detonation,
then 3D \athena\ to model the hydrodynamic interaction of the supernova
ejecta with the donor star, calculating the amount of mass that is
stripped and the entropy deposited in the deep stellar interior by the strong
shock that traverses it. We show that these donor remnants are ejected
with velocities primarily set by their orbital
speeds: $700\text{--}900\ \rm km\ s^{-1}$.
We model the long-term thermal evolution of remnants by
introducing the shock  entropy  into \mesa\ models.
In response to this entropy change, donor remnants expand and brighten
for timescales ranging from $10^6\text{--}10^8$ years, giving ample time for
these runaway stars to be observed in their inflated state before they
 leave the galaxy. Even after surface layers are
stripped, some donors retain enough mass to resume core helium burning
and further delay fading for more than $10^8$ years.

\end{abstract}

\keywords{Runaway stars (1417), High-velocity stars (736), Close
  binary stars (254), Subdwarf stars (2054), White dwarf stars (1799),
  Compact binary stars (283), Supernovae (1668)}

\section{Introduction }

Recent observational and theoretical progress has revived the long considered double 
detonation scenario for thermonuclear supernova
\citep{Nomoto1982,WoosleyWeaver1994,LivneArnett1995}, where a  
shell of accreted He detonates at a strength adequate to detonate the underlying carbon/oxygen (CO)
white dwarf (WD). This totally disrupts the WD, with nucleosynthetic yields 
reflecting the WD core density \citep{Sim2010, Fink2010, Shen2018a, Polin2019}. For thin enough He shells, 
some \citep{Sim2010, Kromer2010, WoosleyKasen2011, Shen2018a, Townsley19} have argued that the diversity of Type~Ia supernovae (SNe) may be explained by the range in the total WD mass 
(C/O core plus He shell) at the time of explosion. Thicker (approximately $>0.03M_\odot$) He shells pose a challenge due to the predicted presence of heavy element ashes from the He detonation on the outer edges of the ejecta  that 
are not seen in normal Type~Ia SNe. Hence, binary scenarios that have lower mass 
He  shells at the time of He detonation, 
such as the Am CVn systems (e.g.~\citealt{Bildsten2007}) or dynamical mass transfer in a merger 
\citep{Guillochon2010, Raskin2012,Pakmor2013,Moll2014,Tanikawa2015} remain the favored scenarios 
for double detonations as the cause of Type~Ia SNe. 

Although a rarer event than Type~Ia SNe, 
the recent observation of \citet{De19} of ZTF18aaqeasu (also referred to as SN~2018byg and ATLAS~18pqq) clearly showed the line blanketing indicative of 
heavy elements on the surface of the ejecta from a more massive He shell detonation that also triggered a core CO detonation. 
\citet{De19}'s interpretation of the SNe spectra and light-curve led them to conclude that
the He shell mass was  $\approx 0.15M_\odot$ on an underlying $0.75M_\odot$ CO WD, 
similar to the prediction of \citet{Bauer17} for the explosive conditions reached in the future for the known galactic binary CD~$-30^\circ 11223$ \citep{Geier13}. 

The binary scenario that naturally creates this explosive environment is a core He burning sdB star of $M_1<0.5M_\odot$  orbiting  a CO WD \citep{IbenTutukov1991}.  Created in a common envelope event, these binaries are found in our galaxy \citep{Geier13, Kupfer2017} and undergo gravitational wave losses on a timescale short enough ($<100$ Myr) to initiate mass transfer while  He is still burning in the core of the sdB star. Reaching contact at an orbital period of  $P_{\rm orb} \approx 30$ minutes, the donor transfers helium to the companion WD  at $\dot M\approx (2\text{--}3)\times 10^{-8} M_\odot\ {\rm y^{-1}}$  \citep{Brooks15,Bauer17,Neunteufel2019}. At these low $\dot M$'s, $\approx 0.05\text{--}0.25M_\odot$ of He accumulates on the WD before it undergoes an unstable thermonuclear flash leading to the triggering  He detonation  \citep{Brooks15, Bauer17}. 

At the time of the He (and presumably CO) detonation, the binary has an orbital period of $10\text{--}15$ minutes. Depending on the WD mass, the whittled down $0.2\text{--}0.35M_\odot$ donors have orbital velocities of  $v_{\rm orb} \approx 600\text{--}900\ \rm km\ s^{-1}$. This velocity is naturally in the range of a set of recently discovered fast moving objects \citep{Geier15,Vennes17,Raddi18mnras,Raddi18apj,Raddi19} that occupy an unusual part of the HR diagram between the main sequence and the WD cooling sequence.  Motivated by this discovery, we explore here the impact of the resulting thermonuclear supernovae on the donor for two specific scenarios that start with low mass He burning star donors. Our 3D simulations yield mass stripping of the donor and the amount of entropy deposited in its deep 
interior due to the shock from the ejecta sweeping across it. We then explore the subsequent longer-timescale expansion and brightening of the donor over thermal timescales and place these shock-heated donors on the HR diagram. Some of our scenarios yield objects close to those observed. 
We do not address  the much higher velocity (up to $\approx2{,}000\ \rm km\ s^{-1}$) objects found by \cite{Shen18}, as the He core burning
donor scenarios cannot reach such compact orbital configurations.

\section{MESA Models of He Star Donors }
\label{sec.donor}

We construct two models of subdwarf He stars that donate material to
WD binary companions using the binary evolution capabilities
of \mesa\ \citep{Paxton11,Paxton13,Paxton15,Paxton18,Paxton19}. 
All \mesa\ models use release
version 10398. Our methods closely follow those of
\cite{Brooks15} and \cite{Bauer17}, including the $^{14}{\rm
  N}(e^-, \nu) {^{14}{\rm C}} ( \alpha, \gamma) {^{18}{\rm O}}$ (NCO)
reaction chain that triggers He shell ignition in the accreting WD.
We employ the rates adopted in \cite{Bauer17} for these reactions.
 
Model~1 is based on the observed system CD~$-30^\circ 11223$
\citep{Geier13}, which consists of a $M_1=0.51\ M_\odot$ He core burning star
with a $M_2=0.762\ M_\odot$ WD companion in a $70.5$ minute orbital
period. Our \mesa\ binary model predicts that gravitational wave
radiation will bring this system into contact in $37\ \rm Myr$
while the donor is still burning helium in its core and the
orbital period is $33$~minutes.  Over  $10\ \rm Myr$, the donor transfers
$0.17\ M_\odot$ of He-rich material to the WD, triggering
a He shell detonation when $P_{\rm orb} = 20.8\ \rm min$. 
Such an event likely causes a SN similar to ZTF~18aaqeasu
(SN~2018byg, \citealt{De19}).
Our binary model for this system is very similar to that of
\cite{Brooks15} with the accretor modeled as in \cite{Bauer17}.  
The only major difference is that for the He core burning donor model
we employ the predictive mixing scheme described in \cite{Paxton18}
for locating the convective boundary during core helium burning,
resulting in a more extended core and $120\ \rm Myr$ core helium burning lifetime.
Changes to the binary mass transfer are negligible, but the predictive
mixing scheme does change the interior profile of the donor at the
moment of the companion explosion. 

Model~2 explores the possibility of a low mass He core burning star that
donates more mass before the WD explodes.   This scenario has  a
more compact orbit with higher orbital velocity at the time of explosion, and  the
potential for the donor to be more impacted by the SN ejecta. 
Model~2 consists of a $M_1=0.46\ M_\odot$ He core burning  star with a $M_2=0.55\ M_\odot$
WD that we initialize with a $70$ minute orbital period. 
After coming into contact at $P_{\rm orb} = 33\ \rm min$, the
donor transfers  $0.23\ M_\odot$ of He-rich material before the
accumulated shell ignites on the accreting WD when
$P_{\rm orb} = 9.4\ \rm min$. 
Due to the donor losing enough mass to fall below $0.3\ M_\odot$,
nuclear burning has ceased in its core \citep{Brooks15}. However, 
the donor's adiabatic response to ongoing mass loss 
 prevents it from contracting and causes
mass transfer to continue even after burning ceases.

Table~\ref{tab:donors} shows the final
properties of the two \mesa\ donor models at the moment of He shell
ignition on the companion WD, which we presume corresponds to a
SN soon thereafter.

\begin{table*}
\caption{Donor properties at the moment the accreted helium shell on
  the companion detonates. Angle brackets denote mass-averaged quantities, 
    such as mass-averaged pressure $\langle p \rangle \equiv \int p \, \mathrm{d}m/M$.}
\label{tab:donors}
\begin{center}
\begin{tabular}{c | lllll | c}
 & Mass ($M_1$) & Radius ($R_1$) & Separation ($a$) & $P_{\rm orb}$ & $v_{\rm orb}$ 
  & Accretor Mass ($M_2$) \\
\hline
Model 1 & $0.344\ M_\odot$ & $0.080\ R_\odot$ & $0.271\ R_\odot$ & $20.8$ min & $691\ \rm km\ s^{-1}$ 
  & $0.927\ M_\odot$ \\ 
Model 2 & $0.233\ M_\odot$ & $0.041\ R_\odot$ & $0.147\ R_\odot$ & $9.4$ min & $882\ \rm km\ s^{-1}$
  & $0.779\ M_\odot$ \\
\hline
  & $\rho_{\rm c }\ [{\rm g\ cm^{-3}}]$ & $T_{\rm c}\ [{\rm K}]$ & $p_{\rm c}\ [{\rm dyne\ cm^{-2}}]$ 
  & $p_{\rm c}/n_{\rm c} k_{\rm B} T_{\rm c}$ & $\langle p \rangle /\langle n k_{\rm B} T\rangle$ \\
\hline
Model 1 & $3.74 \times 10^4$ & $1.10 \times 10^8$ & $2.83 \times 10^{20}$ & 1.23 & 1.21 \\
Model 2 & $6.71 \times 10^4$ & $7.49 \times 10^7$ & $4.55 \times 10^{20}$ & 1.73 & 1.56
\end{tabular}
\end{center}
\end{table*}

For a Roche-lobe filling donor of mass $M_1$ in a system with mass
ratio $q = M_1/M_2 \approx 1/3$, the \cite{Eggleton83} formula gives the ratio of the 
 donor radius $R_1$ to the orbital separation $a$ as
\begin{equation}
\label{eq:eggleton}
\frac{R_1}{a} = \frac{0.49 q^{2/3}}{0.6 q^{2/3} + \ln(1+ q^{1/3})}
\approx 0.3~.
\end{equation}
This shows that values of the ratio $R_1/a$ similar to those seen in
Table~\ref{tab:donors} are generic for the sdB+WD binary evolution
scenario at the time of explosion.  The spatial
velocity of the donor is
\begin{equation}
v_{\rm orb}^2 = \frac{G M_2}{a(1+q)}~.
\end{equation}
Combining this with Equation~\eqref{eq:eggleton}, the radius of the
donor at the time of explosion can be expressed in terms of its orbital
velocity as
\begin{equation}
\label{eq:donor_radius}
R_1 \approx \frac{0.3}{1+q} \frac{G M_2}{v_{\rm orb}^2}
\approx 0.04\ R_\odot \left(\frac{M_2}{M_\odot} \right)
\left( \frac{1000\ \rm km\ s^{-1}}{v_{\rm orb}} \right)^2~.
\end{equation}
Invoking orbital motion to explain high-velocity objects therefore
requires compact radii. For high-velocity objects such as those found by
\cite{Raddi18mnras,Raddi18apj,Raddi19}, their current position on the
HR diagram indicates that they must have expanded in
radius since the explosion for a binary evolution scenario to be a
viable explanation of observed velocities.
For the even higher velocity (up to $\approx2{,}000\ \rm km\ s^{-1}$) objects
of \cite{Shen18}, Equation~\eqref{eq:donor_radius} indicates they must
have had $R\approx 0.01\ R_\odot$ at the moment of explosion, requiring even more
significant inflation to achieve their currently observed states.

 Equation~\eqref{eq:eggleton} also indicates that the fraction
of the solid angle surrounding the exploding WD filled by the donor is
$\pi R_1^2/4\pi a^2 = (R_1/a)^2/4 \approx 0.02$. The total binding
energy of the donor remnants is 
$(2\text{--}4) \times 10^{48}\ \rm erg$. So for an explosion resulting in ejecta with a total
energy of $\approx$$10^{51}\ \rm erg$, the ejecta that intersect the
donor remnant contain sufficient energy to have an impact
on its binding energy and radius, or to unbind significant amounts of
mass from the star, as we now explore.

\section{Estimating the Shock Strength in the Donor} 
\label{sec.expectations} 

  Though none have calculated the specific binary scenario we are exploring here, there has been 
  substantial prior work on SNe ejecta sweeping across nearby companions, both from thermonuclear and core collapse events 
\citep{Wheeler1975,  
  Taam84,Marietta2000,Pan10,Pan12a, Pan12b,Liu13,Pan13,Hirai2014,Hirai2018}. 
  The focus of much of these earlier efforts was on understanding the mass stripped from the donor and the resulting kick. 
Closer to our case, previous studies have examined the interaction of Type~Ia SN
ejecta with He-star companions \citep{Pan10,Pan12b,Pan13,Liu13},
especially with regard to how much mass can be stripped from the
star. \cite{Pan12a,Pan13} also explored post-impact thermal evolution
of the donor using MESA models. As shown by \citet{Pan2014},  deeply injected entropy allowed the 
remaining shocked donor to stay hot (and potentially visible) for hundreds of years after the SNe event.
These, however, were for He-star companions with masses in the range $0.7\text{--}1.2\ M_\odot$, 
much larger than the $\approx0.3\ M_\odot$ remnants that we study here. As we show, the lower masses
of our systems enable a much more prolonged bright phase after the thermonuclear event.

 Though we will investigate mass loss and kicks from the ejecta momentum, we want to
 emphasize here the thermal impact on the donor of the shock that traverses its core.
As the entropy jump associated with the deeply penetrating shock wave depends on the shock pressure versus that in the  
ambient star, we start by estimating the ejecta pressure at the location of the donor, 
\begin{equation}
p_{\rm ej}\approx \frac{M_2 v_{\rm ej}^2/2}{4\pi a^3/3},
\label{eq:pejest}
\end{equation}
where $v_{\rm ej}$ is the mass averaged ejecta velocity. When the donor is well characterized 
as an $n=3/2$ polytrope, it's central pressure is $p_{\rm c,p}=0.77 GM_1^2/R_1^4$. Assuming that the donor
is also Roche-lobe filling then yields the ratio
\begin{equation}
  \frac{p_{\rm ej}}{p_{\rm c,p}} \approx \frac{7\times 10^{-3}}{q^{2/3}(1+q)^{7/3}} \left(\frac{v_{\rm ej}}{v_{\rm orb}}\right)^2.
\label{eq:pratios}
\end{equation}
This equation's prime value is in the scaling that indicates a much larger impact on the entropy for the widest Roche-lobe 
filling binaries. An excellent case of this is shown by  \citet{Taam84} where a much wider binary polytropic 
companion suffers a very large central pressure
perturbation. It is also evident in \citet{Marietta2000}'s work on a model of a near solar analog star (their model HCV) in a 
$\approx 9.75 $ hour Roche-lobe filling orbit around a
$M_2=1.378M_\odot$ WD. In their case, $v_{\rm orb}\approx 230 \ {\rm km \ s^{-1}}$, 
and Equation~\eqref{eq:pratios} predicts $p_{\rm ej}/p_{\rm c,p}\approx 2$ for their ejecta model.
This was explicitly noted by \citet{Marietta2000} as they diagnosed
the outcome of their simulation. The later simulations by
\citet{Pan12a,Pan12b,Pan13} also exhibited substantial interior
thermal perturbations. 

If our donors were simple $n=3/2$ polytropes, then Equation~\eqref{eq:pratios} would imply that 
$p_{\rm ej}/p_{\rm c,p}\approx 0.3$ for $v_{\rm ej}=5{,}000 \ {\rm km \ s^{-1}}$ certainly indicating 
the need to perform the rigorous 3D calculation that follows. However, having undergone substantial burning that modifies their compositions throughout, as well as having internal entropy gradients, our donors are far from polytropes. Despite that, we can use 
the values in Table~\ref{tab:donors} to make a few preliminary estimates. Model 1 has a central pressure about a factor of ten larger than 
an $n=3/2$ polytrope, whereas Model 2 is about a factor of three higher in central pressure than a polytrope. The resulting values 
of $p_{\rm ej}/p_{\rm c}$ for a $10^{51}$ erg explosion are 0.1 for Model 1 and 0.5 for Model 2. It's also important to note that 
the volume averaged pressure in a stellar model is $\langle p\rangle=-2E_{\rm tot}/3V$ where $V$ is the volume of the star
and $E_{\rm tot}$ is its total energy. Hence for a $10^{51}$ erg explosion,  Model 1 ($E_{\rm tot}=-4.2\times 10^{48} \ {\rm erg}$)
has  $p_{\rm ej}/\langle p\rangle \approx 9$ , while Model 2 ($E_{\rm tot}=-2.6\times 10^{48}\  {\rm erg}$) has 
 $p_{\rm ej}/\langle p\rangle \approx 12$. Hence, for both cases a very large part of the 
 donor's volume (though maybe not its mass) will undergo a strong shock.  As we will see, this leads to mass loss 
 at different levels.

\section{Ejecta-Donor Interaction Computations }
\label{sec.impact}

The density and pressure profiles of the two \mesa{} models are used
to initialize 3D \athena{} models on $256^3$ grids. Model~1 has an
initial diameter of $46$ cells on its grid, while Model~2 has a
diameter of $45$ cells. As the donor stars are nearly non-degenerate
(as indicated in Table 1) \athena{} is run with $\Gamma = 5/3$ ideal
hydrodynamics and with self gravity based on the fast Fourier
transform. The initial stars are allowed to settle into numerical
hydrostatic equilibrium on these new grids for $30$ characteristic
dynamical times $t_\mathrm{dyn} = \sqrt{5 / 8 \pi G \rho_\mathrm{c}}$
(totaling $270\ \mathrm{s}$ and $200\ \mathrm{s}$, respectively)
before interacting with the modeled ejecta. Though the
resolution used is not sufficient to fully resolve the tenuous
stellar atmosphere, the stars quickly find this new equilibrium,
which is very similar to the \mesa{} model throughout the interior
(see Section~\ref{sec.post} for more detail).

For the ejecta, we use the model presented in \citet{Kasen2010} with power-law slopes $\delta = 1$ and $n = 10$, truncated to have velocity less than $20{,}000\ \mathrm{km\ s^{-1}}$. The ejecta mass is set to be the accretor mass at explosion, and the ejecta kinetic energy is set to be $0.2$, $0.5$, $0.7$, or $1.0 \times 10^{51}\ \mathrm{erg}$. These kinetic energies are chosen to explore the range of possible outcomes.  The corresponding mass-averaged ejecta velocities span the ranges $v_\mathrm{ej} = 4{,}330\text{--}9{,}740\ \mathrm{km\ s^{-1}}$ (Model~1) and $v_\mathrm{ej} = 4{,}720\text{--}10{,}600\ \mathrm{km\ s^{-1}}$ (Model~2). To account for the losses from adiabatic expansion of the ejecta's internal energy from the time of explosion, we set the internal energy density of the ejecta to be proportional to $1/t$ times the kinetic energy density when it enters the grid at time $t$ after explosion. The normalization is such that $10{,}000\ \mathrm{km\ s^{-1}}$ ejecta will have an internal energy $7\text{--}9\%$ of its kinetic energy when it reaches the donor. When incorporating the ejecta into the 3D simulation, we shift the velocities to account for the relative orbital motion between the donor and accretor.

\begin{figure*}
  \centering
  \includegraphics{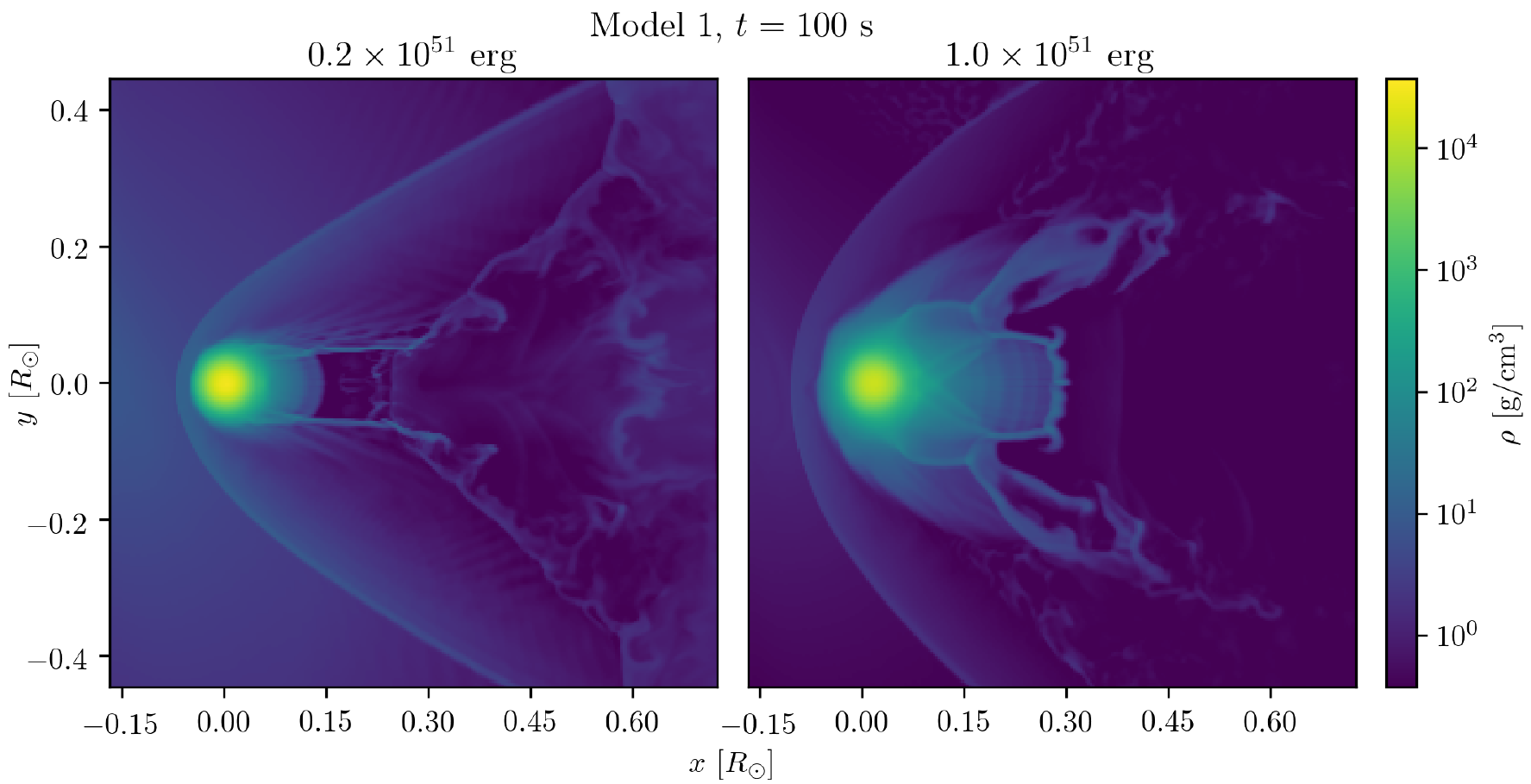}
  \caption{Density in two simulations of ejecta interacting with Model~1. The lowest ejecta energy is on the left, with the highest on the right. Both snapshots are taken $100\ \mathrm{s}$ after explosion. The online animation shows time evolution of this density over the entire simulation. \label{fig:3d_slice_1}}
\end{figure*}

\begin{figure*}
  \centering
  \includegraphics{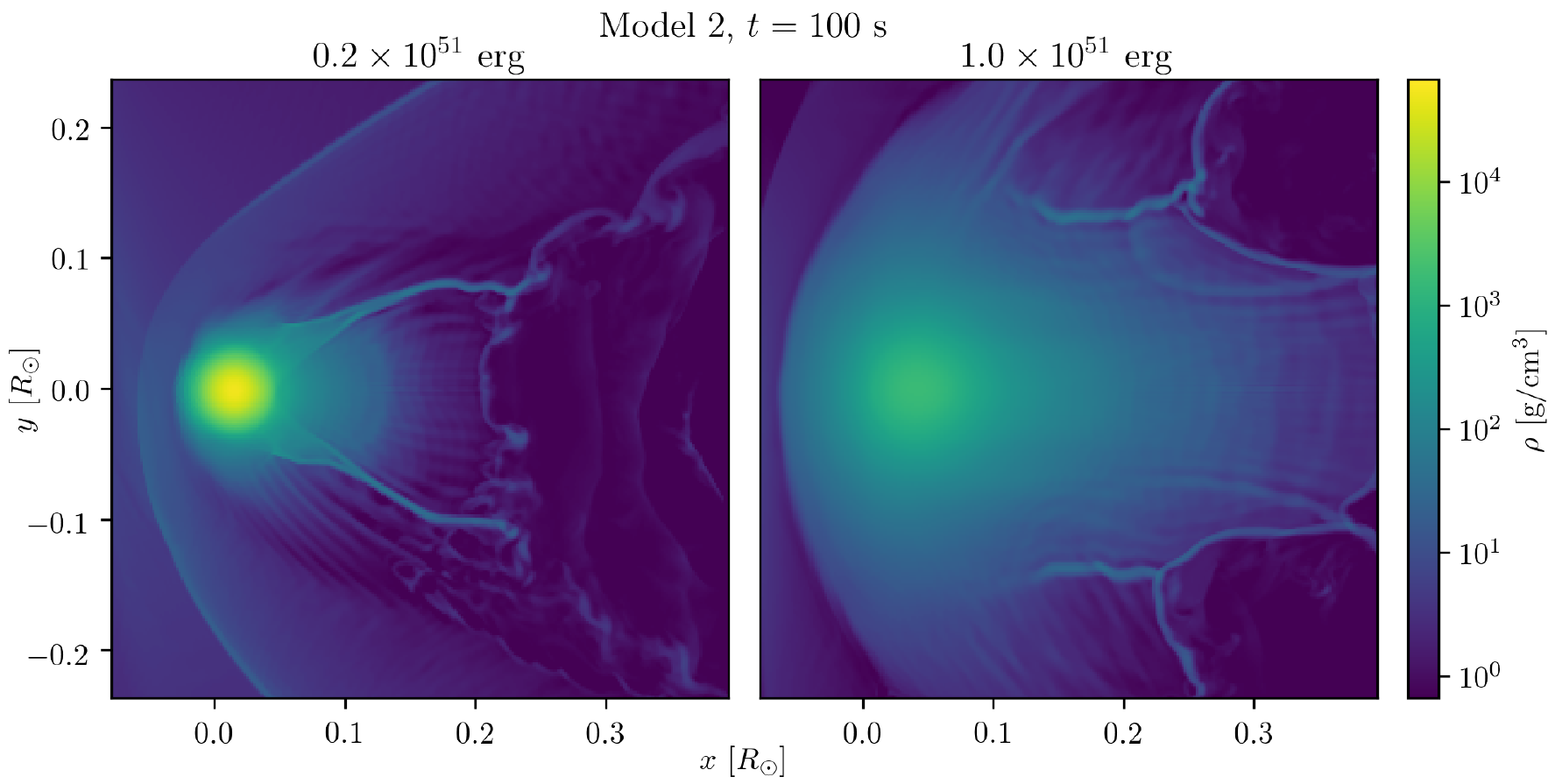}
  \caption{Density in two simulations of ejecta interacting with Model~2. The lowest ejecta energy is on the left, with the highest on the right. Both snapshots are taken $100\ \mathrm{s}$ after explosion. The online animation shows time evolution of this density over the entire simulation. \label{fig:3d_slice_2}}
\end{figure*}

Figure~\ref{fig:3d_slice_1} shows the density in the orbital plane for two of the 3D \athena{} computations with Model~1, each $100\ \mathrm{s}$ after explosion. The WD accretor is located off the domain, $0.2706\ R_\odot$ to the left of the origin, with the donor's orbital velocity in the positive $y$-direction. The left panel shows the $0.2 \times 10^{51}\ \mathrm{erg}$ case, while the right panel shows the $1.0 \times 10^{51}\ \mathrm{erg}$ case. Figure~\ref{fig:3d_slice_2} shows the same density slices for Model~2, also $100\ \mathrm{s}$ after explosion. In this case, the WD accretor is $0.147\ R_\odot$ to the left of the origin.

\begin{figure*}
 \centering
 \includegraphics{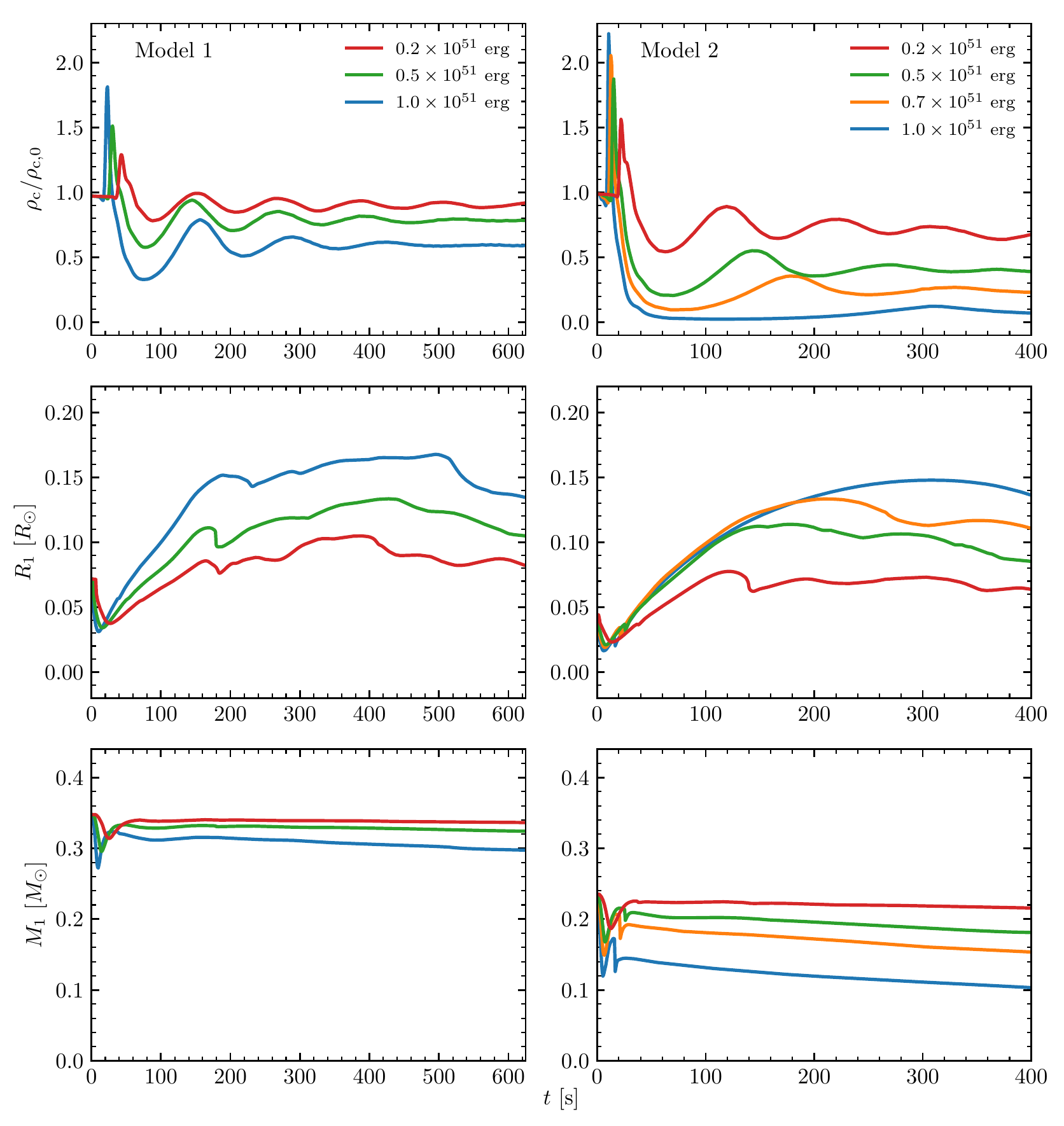}
 \caption{Central density, radius, and bound mass of the two donors as functions of time after explosion. Larger kinetic energies in the ejecta cause stronger oscillations and increase the radii more, while also stripping more mass. \label{fig:3d_time}}
\end{figure*}

As noted earlier, when hit with the ejecta, a strong shock passes through the donor, inducing a series of pulsations that decay over many dynamical times. This can be seen in the top panels of Figure~\ref{fig:3d_time}, which show the time evolution of the central density normalized by the central density of the \mesa{} model given to \athena. The central pressure is qualitatively the same. The initial spikes in the central density increase with explosion energy, as expected.

In the 3D computations we  define the extent of the star at any time using the following approach.  Given the center of mass, we construct spherically averaged radial profiles of the Bernoulli parameter
\begin{equation}
  \mathrm{Be} \equiv \frac{1}{2} v^2 + \frac{\Gamma}{\Gamma-1} \paren[\bigg]{\frac{p}{\rho}} + \Phi,
\end{equation}
where $v$ is the velocity measured relative to the star's bulk motion and $\Phi$ is the gravitational potential relative to $0$ at infinity. We take the edge of the star to be the innermost radius where the Bernoulli parameter vanishes. The results do not change much if we use total specific energy $\mathrm{Be} - p/\rho$ instead, though the Bernoulli parameter properly accounts for a fluid element being able to reach infinity using not just its kinetic energy and, via cooling, its internal energy, but also using its pressure via expansion into vacuum. With this definition, we can measure the radial extent and total bound mass of the donor as a function of time as plotted in Figure~\ref{fig:3d_time}. While these radii do sometimes exceed the initial distance between the donor's center and the edge of the grid, kicks provided by the ejecta move the donor sufficiently far from the edge for the bound material to never leave the grid in any case.

\begin{figure}
 \centering
 \includegraphics{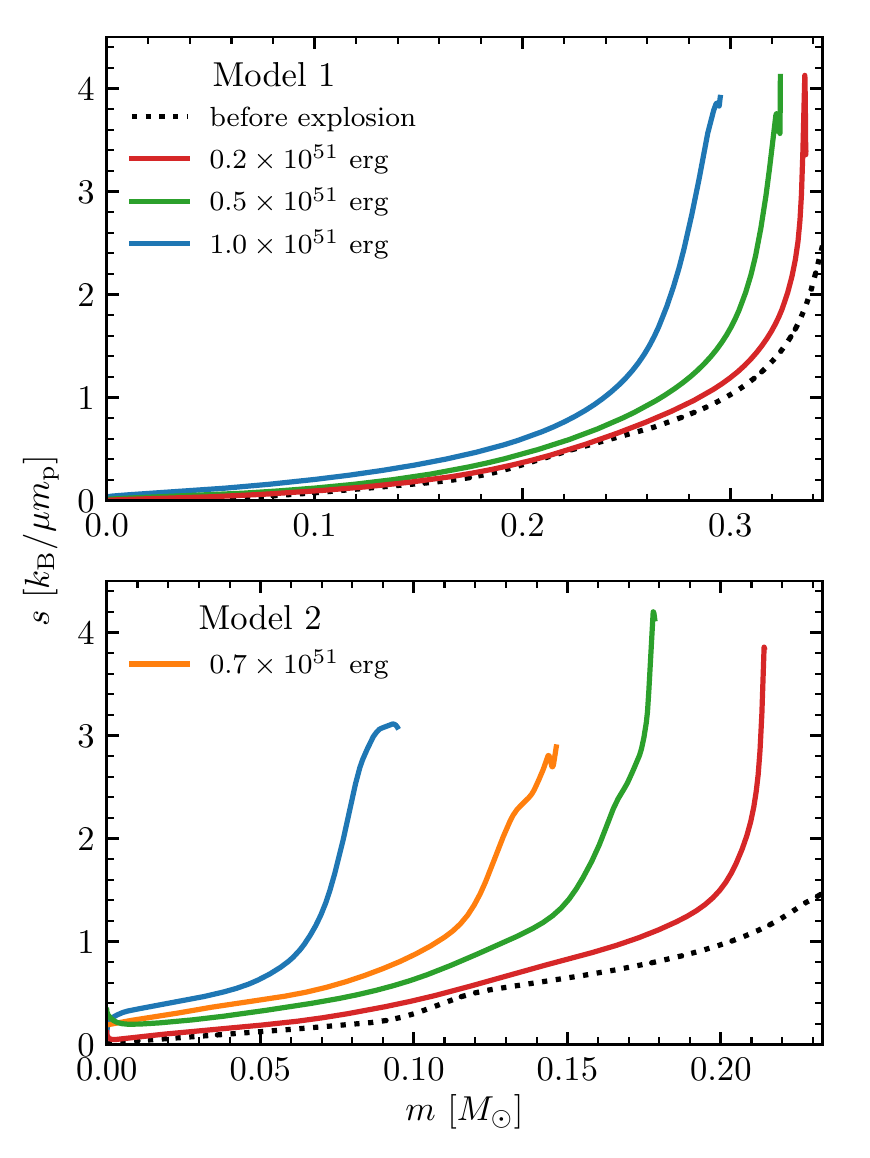}
 \caption{Entropy profiles for the two models, before and after interacting with the supernova. \label{fig:3d_entropy}}
\end{figure}

The donors' outer radii dramatically increase after interacting with the ejecta, especially in the higher-energy explosions. This occurs for two reasons. The first reason is the expected hydrostatic expansion of a star due to rapid mass loss, while the second reason is the increase in entropy deep in the star due to the shock wave traversal. As we show later, the  amount and location of this deposition of heat will determine the appearance of the star at much later times. We show the effect of the shock wave traversal by plotting the spherically averaged profiles of entropy per unit mass relative to the initial central value,
\begin{equation}
  s \equiv \frac{k_\mathrm{B}}{(\Gamma-1) \mu m_\mathrm{p}} \log\paren[\bigg]{\frac{p}{\rho^\Gamma} \frac{\rho_{\mathrm{c},0}^\Gamma}{p_{\mathrm{c},0}}},
\end{equation}
where $k_\mathrm{B}$ is Boltzmann's constant, $\mu$ is the mean molecular weight, and $m_\mathrm{p}$ is the baryon mass. Figure~\ref{fig:3d_entropy} shows the initial entropy profiles, as well as the profiles at the end of the 3D simulations, $70$ dynamical times after the explosion. The substantial entropy gradient in the initial models helps to explain why our polytropic estimates for central pressures were not accurate. The large increase in entropy in the outermost layers reflect the much stronger shocks that can be achieved at the lower pressures there.

\floattable
\begin{deluxetable}{cDDDDDDDD}
  \tablecaption{Summary of 3D modeling. \label{tab:3d_properties}}
  \tablewidth{0pt}
  \tablehead{\colhead{Model} & \multicolumn{2}{c}{$M_\mathrm{f}$ [$M_\odot$]} & \multicolumn{2}{c}{$\Delta s_\mathrm{ave}$ [$k_\mathrm{B}/\mu m_\mathrm{p}$]} & \multicolumn{2}{c}{$\Delta s_\mathrm{c}$ [$k_\mathrm{B}/\mu m_\mathrm{p}$]} & \multicolumn{2}{c}{$\rho_\mathrm{c,f}/\rho_\mathrm{c,i}$} & \multicolumn{2}{c}{$p_\mathrm{c,f}/p_\mathrm{c,i}$} & \multicolumn{2}{c}{$v_\mathrm{kick}$ [$\mathrm{km\ s^{-1}}$]} & \multicolumn{2}{c}{$v_\mathrm{f}$ [$\mathrm{km\ s^{-1}}$]} & \multicolumn{2}{c}{$4\pi I_\mathrm{kick}/\Omega_xI_\mathrm{ej}$}}
  \decimals
  \startdata
  1:0.2 & 0.336  & 0.077 & 0.00  & 0.95  & 0.91  & 52  & 680 & 0.19 \\
  1:0.5 & 0.324  & 0.21  & 0.012 & 0.80  & 0.70  & 120 & 700 & 0.28 \\
  1:1.0 & 0.297  & 0.40  & 0.034 & 0.61  & 0.45  & 180 & 710 & 0.28 \\
  2:0.2 & 0.214  & 0.20  & 0.11  & 0.70  & 0.60  & 150 & 880 & 0.44 \\
  2:0.5 & 0.178  & 0.50  & 0.33  & 0.37  & 0.24  & 270 & 920 & 0.42 \\
  2:0.7 & 0.150  & 0.63  & 0.19  & 0.25  & 0.11  & 310 & 930 & 0.34 \\
  2:1.0 & 0.0952 & 0.93  & 0.17  & 0.076 & 0.015 & 330 & 940 & 0.19 \\
  \enddata
  \tablecomments{Models are labeled according to donor model number and the kinetic energy of the explosion in units of $10^{51}\ \mathrm{erg}$.}
\end{deluxetable}

Table~\ref{tab:3d_properties} summarizes the outcomes of the seven 3D hydrodynamical models, labeled by the initial 1D model number and the kinetic energy of the explosion. The final bound remnant always has a mass less than the initial donor mass. The density floor used in the modeling is $10^{-5}$ the initial central density. As a result this floor material will only have a mass $0.04\%$ ($0.01\%$) that of the initial Model~1 (Model~2) donor in an equal volume, and it will comprise a mass of $13\%$ ($5\%$) that of the initial donor over the entire simulation volume.

We define the change in average entropy per unit mass using the initial  and final entropy profiles:
\begin{equation}
  \Delta s_\mathrm{ave} = \frac{1}{M_\mathrm{f}} \int_0^{M_\mathrm{f}} (s_\mathrm{f} - s_\mathrm{i})\ \mathrm{d}m,
\end{equation}
This increases with increasing explosion energy, as expected. We also report $\Delta s_\mathrm{c}$, the change in entropy per unit mass at the center of the donor, as well as the fractional changes in central density and pressure.

The explosion delivers an impulse to the donor over a relatively short time, after which the donor coasts at a well-defined kick velocity $v_\mathrm{kick}$ for the remainder of the simulation. These velocities, calculated in the initial rest frame of the donor, are dominated by the component in the $x$-direction in the sense of Figures~\ref{fig:3d_slice_1} and~\ref{fig:3d_slice_2}. Due to the drag on the donor moving through the ejecta (from its orbital motion), there is also a slight negative component in the $y$-direction, which becomes more important at lower ejecta velocities, but it only ranges over $3\text{--}13\ \mathrm{km\ s^{-1}}$ across all simulations. Though $v_\mathrm{kick}$ can be a nonnegligible fraction of $v_\mathrm{orb}$, the two are largely orthogonal and the latter dominates the final velocity $v_\mathrm{f}$ of the donor relative to the binary barycenter. In fact, in some cases the small drag in the negative $y$-direction leads to a reduced final barycentric velocity, despite the kick in the $x$-direction. These velocities are reported in Table~\ref{tab:3d_properties}.

The final column of Table~\ref{tab:3d_properties} measures the fraction of ejecta momentum intercepting the donor that contributes to the final velocity. That is, we measure the kick impulse
\begin{equation}
  I_\mathrm{kick} = M_\mathrm{f} v_\mathrm{kick}^x,
\end{equation}
and compare it to the total impulse in the ejecta model
\begin{equation}
  I_\mathrm{ej} = \int_0^{M_\mathrm{ej}} v_\mathrm{ej}\ \mathrm{d}m_\mathrm{ej}.
\end{equation}
The latter must be scaled by the cross section to the explosion presented by the donor, with each line of sight weighted by the ratio of the ejecta $x$-momentum to the total ejecta momentum. The resulting scale factor is $\Omega_x / 4\pi$, where $\Omega_x = \pi R^2 / a^2$. The resulting values are less than unity, indicating some $x$-momentum in the ejecta is deflected around the star. We expect this to be the case, given that the ejecta has a finite Mach number and forms a visible bow shock when interacting with the donor. For Model~2, much of the momentum lost by the ejecta in the highest energy explosions goes toward accelerating ultimately unbound material, resulting in low intercepted momentum fractions for the bound remnant. In all of our cases, we find that $4\pi I_\mathrm{kick}/\Omega_xI_\mathrm{ej}$ is in the range of $1/3$ noted by \citet{Hirai2018}.

While the modeling here only considers ejecta from a nonrotating accretor, little would change with the addition of rotation. The accretor surface breakup velocities at the time of explosion are $4300\ \mathrm{km\ s^{-1}}$ and $3400\ \mathrm{km\ s^{-1}}$ for Models~1 and~2, respectively. Even if all the ejecta were moving this rapidly in the tangential direction at explosion, the tangential velocities at the location of the donor would be reduced by the ratio of the separation to the accretor radius, resulting in $150\ \mathrm{km\ s^{-1}}$ and $300\ \mathrm{km\ s^{-1}}$. These values are $16\%$ and $26\%$ of the $y$-velocities already seen due to the orbital motion. This would slightly modify the $3\text{--}13\ \mathrm{km\ s^{-1}}$ impact on the donor velocities due to drag, and in fact prograde motion would operate to reduce this already small drag effect.

\section{Post-Interaction Evolution with MESA }
\label{sec.post}

After the oscillations in the bound remnants have died away (see
Figure~\ref{fig:3d_time}), we record shellular averages of the
Lagrangian change $\Delta \ln (p/\rho^\Gamma)$ from the \athena{}
models for fluid elements in the remaining stellar interior.
The profile for the local entropy change is then given by
\begin{equation}
\label{eq:deltaS}
\Delta s
\equiv \frac 3 2 \frac{k_{\rm B}}{\mu m_{\rm p}} 
\Delta \ln\left(\frac{p}{\rho^\Gamma} \right),
\end{equation}
which we inject over an arbitrary time interval $\Delta t$ into the \mesa\ donor
models as a local heating term
\begin{equation}
\label{eq:heat}
\epsilon_{\rm heat} = T \frac{\Delta s}{\Delta t} ~,
\end{equation}
where $T$ is the local temperature.
During the entropy injection phase, we set the model timesteps to be
about one second over a typical time interval $\Delta
t = 100\ \rm s$, after which heating shuts off and we begin tracking the
subsequent thermal evolution. The $\Delta t\approx 100\ \rm s$ is
long enough so that the star can hydrostatically readjust. 
 The temperature change due to this heating is
\begin{equation}
\label{eq:deltaT}
\frac{\Delta T}{T} = \frac{\Delta s}{c_V}~,
\end{equation}
where $c_V \approx (3/2)k_{\rm B}/\mu m_{\rm p}$ is the specific heat
capacity at constant volume. This implies that 
 profiles of relative temperature change can be inferred 
 from the differences between the 
dashed and solid curves in Figure~\ref{fig:3d_entropy}.

After injecting these entropy changes over the interior mass that
remains bound in each model,
we also strip mass from the surface of the \mesa\ model to match the
final bound mass seen in the \athena\ runs
(Table~\ref{tab:3d_properties}). We remove mass using the
  \mesa\ option \texttt{relax\_mass} with a mass loss rate of
  $10^{-4}\ M_\odot\ \rm yr^{-1}$. This relaxation procedure removes
  mass adiabatically and performs pseudo-evolution to reconverge to
  hydrostatic equilibrium while suppressing any composition changes
  due to mixing or nuclear burning.
  The timescales for mass stripping are on the order of
  $10^3\ \rm yr$ or less, so transient behavior on shorter timescales
  than this due to thermal readjustment near the surface in the \mesa\
  models should be ignored. The remainder of this section focuses
  on the structure changes due to entropy injection in deeper layers,
  where we show that thermal adjustment timescales are longer than the
  mass stripping timescale. The entropy profile stays nearly constant
  in these layers during mass stripping.

\begin{figure}
\begin{center}
\includegraphics[width=\apjcolwidth]{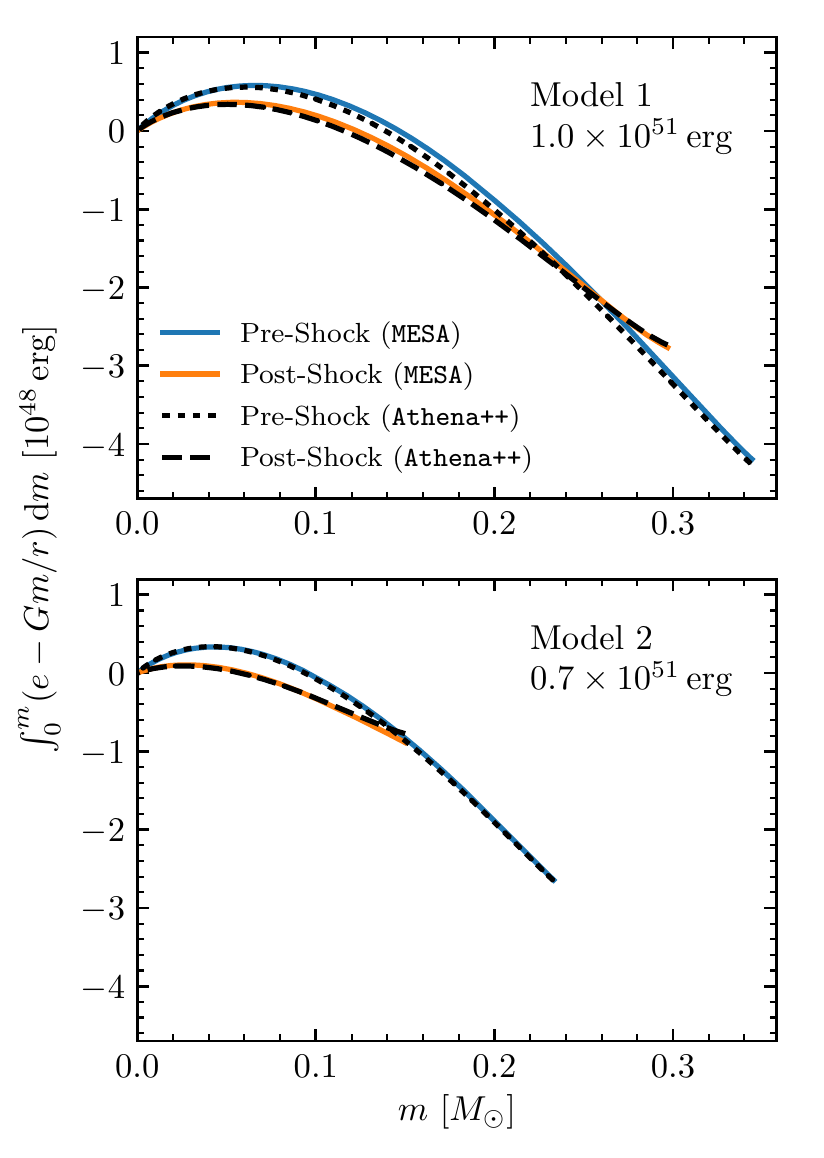}
\caption{Integrated energy profiles (thermal $+$ potential) showing
  the overall change in binding energy after the shock from the
  supernova ejecta deposits entropy and strips mass from the surface.}
\label{fig:binding}
\end{center}
\end{figure}

Figure~\ref{fig:binding} shows the change in binding
energy from this procedure, and compares the energy profiles in the
\mesa\ model to the energy profiles from the \athena\ run.
The ``pre-shock'' \athena{} profiles are shown after the models
have settled into hydrostatic equilibrium on the \athena{} grid,
demonstrating that although there are small changes
due to different equation-of-state treatments when initializing
the \mesa{} progenitor model into the \athena{} simulation, the
differences are small enough that we can still resolve the changes in
the bulk structure due to the shock that traverses the donor star.
The evident agreement of post-shock models in Figure~\ref{fig:binding}
verifies that our procedure for adjusting the \mesa\ model
accurately captures the change in binding energy from the ejecta
interaction modeled with \athena{}.

\begin{figure}
\begin{center}
\includegraphics[width=\apjcolwidth,height=4.73in]{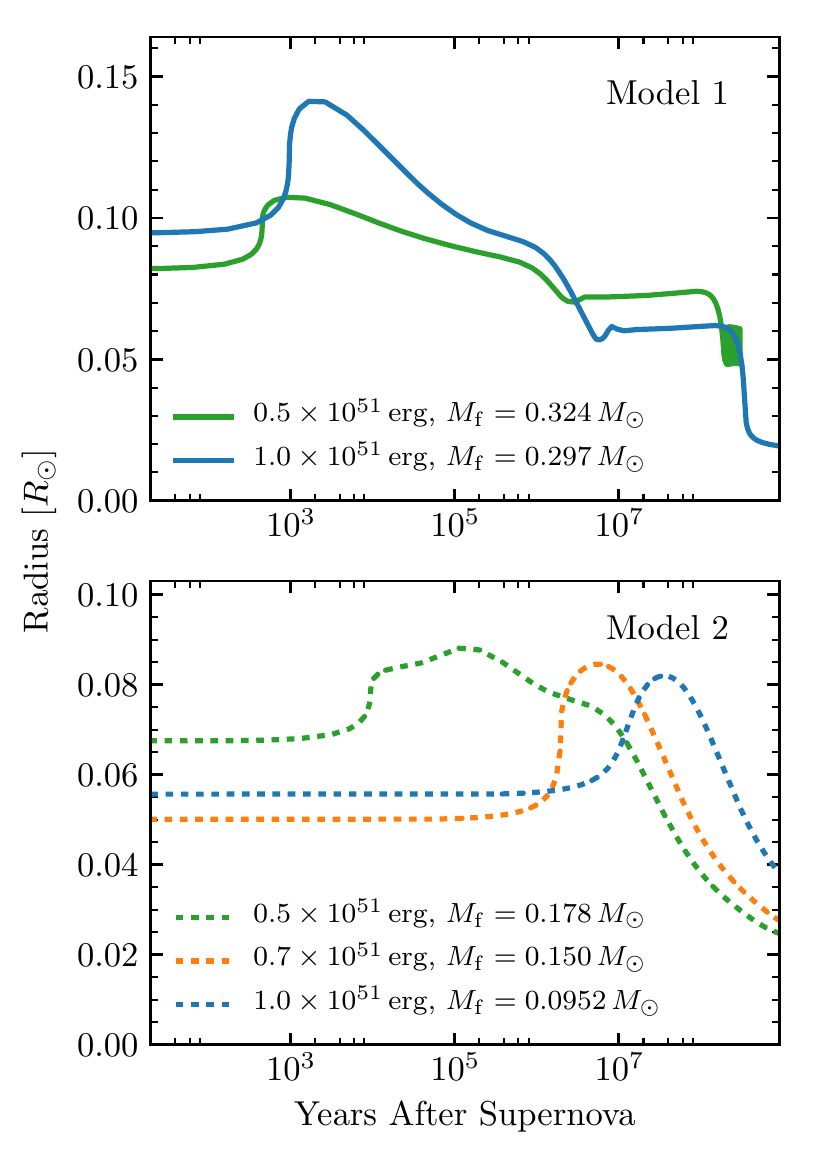}
\caption{Long-term radius evolution of \mesa\ models
 after introducing the entropy change from ejecta interaction in the
 \athena\ models. The labels for each line correspond to the total
 kinetic energy of the ejecta in the \athena\ run.
 {\it Note:} We omit the two cases with the lowest energy explosions
 ($0.2\times 10^{51}\ \rm erg$) from this and subsequent figures to
 improve clarity and because the lowest ejecta energies have only
 superficial impact for short timescales.
}
\label{fig:radius}
\end{center}
\end{figure}

\begin{figure}
\begin{center}
\includegraphics[width=\apjcolwidth,height=4.73in]{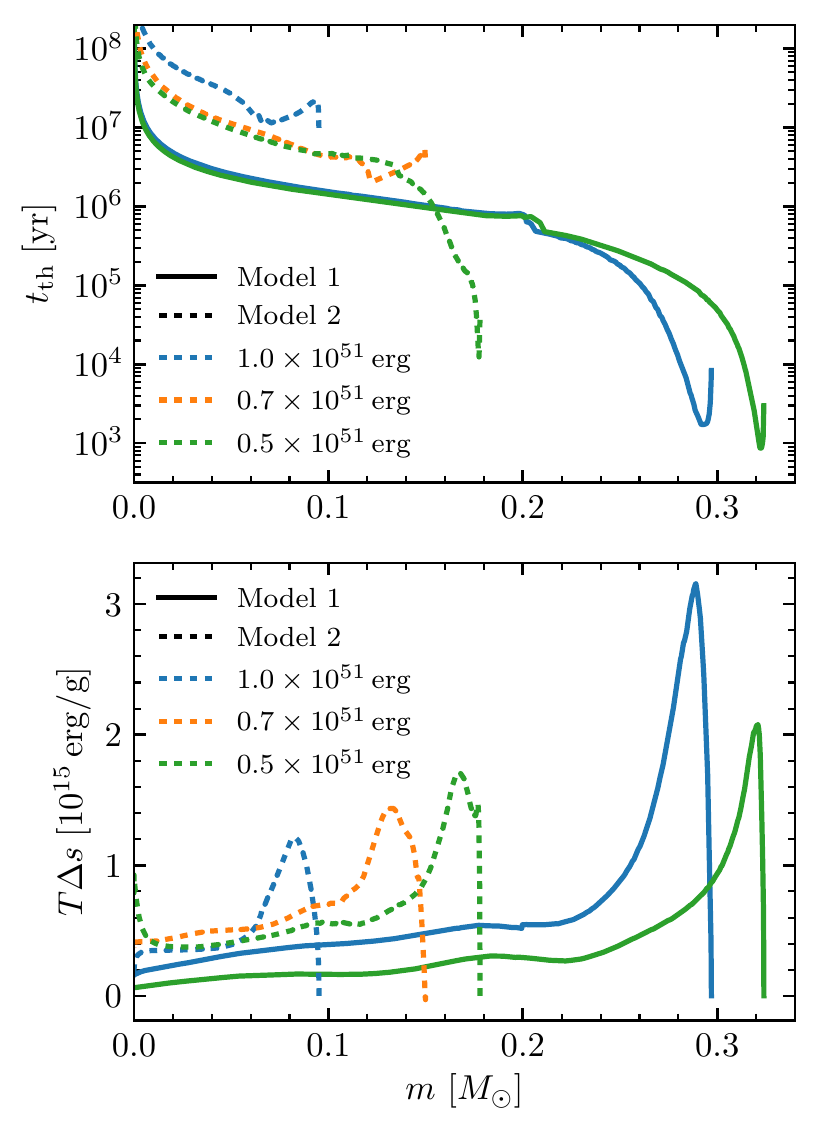}
\caption{{\it Upper Panel:} Profiles of the local heat diffusion timescale
  (Equation~\ref{eq:tth}) in the \mesa\ models for the remnants
  immediately following entropy injection. {\it Lower Panel:}
  Profiles of the total heating introduced into the \mesa\ remnant
  models by the procedure described in Equations~\eqref{eq:deltaS}
  and~\eqref{eq:heat}.
}
\label{fig:timescale}
\end{center}
\end{figure}

The models will then thermally respond to the new entropy profiles.  The heating of the interior results in inflation to larger radii, analogous to the thermal wave described by
\cite{Zhang19} for WDs with entropy deposited in the
interior by nuclear heating. 
Figure~\ref{fig:radius} shows the long-term radius evolution for models~1 and~2.
The timescale for the initial radius expansion and brightening is set by the
local heat diffusion timescale at the location of peak heating,
\begin{equation}
\label{eq:tth}
t_{\rm th} = \frac{H^2}{D_{\rm th}}~,
\end{equation}
where $H=P/\rho g$ is the local pressure scale height and
${D_{\rm th} = 4 ac T^3/3 \kappa \rho^2 c_P}$ is the thermal diffusion
coefficient. For a Kramers opacity ($\kappa \propto T^{-3.5}$), this
timescale depends strongly on temperature: $t_{\rm th} \propto
T^{-6.5}$.

Figure~\ref{fig:timescale} shows the profiles for total
heating in the interior of each remnant model along with interior
profiles of $t_{\rm th}$ at the end of the heat injection phase.
The timescale for radius expansion  seen in
Figure~\ref{fig:radius} corresponds to the value of $t_{\rm th}$ at
the location of peak heating seen in Figure~\ref{fig:timescale}.
Note that even though the initial thermal diffusion timescale in
layers exterior to the peak heating location can be longer, the
thermal wave propagating through them heats and adjusts the structure
of these layers as it reaches them, significantly reducing the
timescale for heat transport \citep{Zhang19}.
It is therefore only the local thermal time for the wave to begin
propagating that sets the timescale for its emergence from the star.
Note that although Table~\ref{tab:donors} shows that the core
  pressure of model~2 is somewhat non-ideal due to the onset of electron
  degeneracy, Figures~\ref{fig:3d_entropy} and~\ref{fig:timescale}
  show that the most important layers for entropy and heat deposition
  lie toward the surface of the star, where conditions are much closer
  to ideal gas. Effects of a non-ideal equation-of-state in the
  \athena{} models for donor stars will be a subject of future work,
  but we do not expect significant changes for the subdwarf donor
  stars presented in this work.

The expansion and increased luminosity persist over the 
Kelvin--Helmholtz timescale $t_{\rm KH} \sim GM^2/RL$, which ranges from
$10^5\text{--}10^8$ years depending on the luminosity $L$ after the thermal
wave has reached the surface of the donor remnant. The trend is for the
overall duration of the brightening event to be shorter for more
massive remnants due to higher peak luminosity. This trend is also
consistent with the results of \cite{Pan13}, whose
models exhibit similar brightening events but with much shorter duration
($10\text{--}100$~years) for their more massive ($0.7\text{--}1.2\ M_\odot$)
He-star companion models that brightened to $10^3\text{--}10^4\ L_\odot$.

The entropy deposited in the core of
model~1 initially reduces the density and temperature there,
temporarily halting core helium burning, but because
 $M_2>0.3\ M_\odot$, these models eventually contract enough to 
resume core He burning and achieve a fixed radius and luminosity
lasting $\approx100\ \rm Myr$, thus making an unusually low mass He core burning star. For model~2, the timescale is much longer
(up to $10^8$ years) due to the much lower remnant luminosity.
After their initial phase of radius expansion,
the remnants from model~2 cool and contract as
low-mass WDs incapable of any further nuclear burning.

\begin{figure}
\begin{center}
\includegraphics[width=\apjcolwidth]{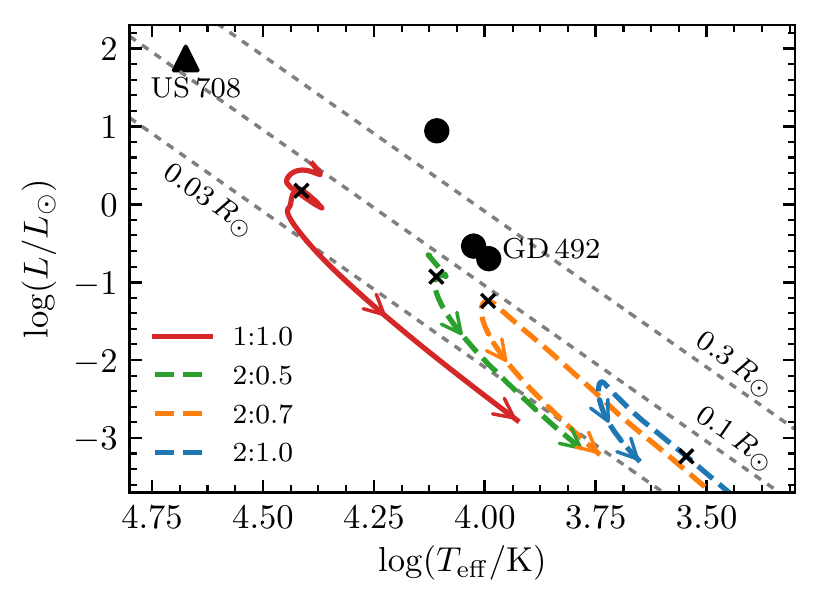}
\caption{HR diagram. Tracks are labeled as the corresponding models in
  Table~\ref{tab:3d_properties} and begin $10^5$ years after the companion
  explosion to avoid showing early portions of evolution that are
  unlikely to be observed. A black x marks the point $10^7$ years
  after the companion explosion on each track. The three circles show
  high-velocity objects of \cite{Raddi19}, including GD~492.}
\label{fig:HR}
\end{center}
\end{figure}

Figure~\ref{fig:HR} compares the model HR diagram tracks to
a few high-velocity objects. Notably, with the right
ejecta energy, model~2 can evolve near the locations on the HR diagram
for GD~492 (LP~40--365) and J1603--6613
\citep{Vennes17,Raddi18mnras,Raddi18apj,Raddi19}, and remain there for
the $10^6\text{--}10^7$ years expected from their kinematics and location in
the galaxy. The ejection velocities for our models are somewhat higher
than the $\approx600\ \rm km\ s^{-1}$ inferred by
\cite{Raddi18mnras,Raddi19} for these objects when accounting for the
rotation velocity of the galactic disk.

We also include an estimated point for US~708 using the values
of $T_{\rm eff}$ and $\log g$ given by \cite{Geier15}, where $L = 4
\pi R^2 \sigma_{\rm SB} T_{\rm eff}^4$ and we estimate $R$ using
the measured $\log g$ and assuming that $M\approx 0.3\ M_\odot$.
Our track for model~1 suggests that a more massive object is needed to
achieve the higher temperature and luminosity of
US~708, though in this case it may be difficult for the binary system
to achieve a compact enough orbit to explain the high velocity of
US~708. However, \cite{Brown15} have noted that the higher velocity of
US~708 could be explained if it originated from a binary in the
stellar halo traveling at $\approx400\ \rm km\ s^{-1}$ in the same direction
that the remnant is ejected from the binary.

\section{Conclusions and Future Work}
\label{sec.conclusion}

\begin{table*}
\caption{Rotational properties of donor remnants $10$ Myr after explosion.}
\label{tab:rotation}
\begin{center}
\begin{tabular}{c | c | ccc | ccc}
Model  & & \multicolumn{3}{c}{Case 1 (see text)} & \multicolumn{3}{c}{Case 2 (see text)} \\
\hline
 & $P_{\rm rot,0}$ [min] 
 & $P_{\rm rot}$ [min] & $\Omega_{\rm rot}/\Omega_{\rm crit}$ & $v_{\rm rot}$ [$\rm km\ s^{-1}$]
 & $P_{\rm rot}$ [min] & $\Omega_{\rm rot}/\Omega_{\rm crit}$ & $v_{\rm rot}$ [$\rm km\ s^{-1}$] \\
\hline
1:0.5 & $20.8$ & $61$  & $0.094$ & $86$ & $20$ & $0.28$ & $261$ \\
1:1.0 & $20.8$ & $52$  & $0.080$ & $80$ & $17$ & $0.25$ & $249$ \\
2:0.5 & $9.4$  & $61$  & $0.093$ & $71$ & $14$ & $0.40$ & $301$ \\
2:0.7 & $9.4$  & $161$ & $0.064$ & $38$ & $28$ & $0.37$ & $219$ \\
2:1.0 & $9.4$  & $230$ & $0.048$ & $24$ & $62$ & $0.18$ & $88$ \\
\end{tabular}
\end{center}
\end{table*}

We have modeled WD+sdB binary star systems and shown that the donor
sdB stars can be expected to become unusual runaway stars with
velocities of $600\text{--}900\ \rm km\ s^{-1}$.
The onset of mass transfer in these binary systems 
occurs at $P_{\rm orb} \approx 30\ \rm min$,
when a $0.46\ M_\odot$ star fills the Roche-lobe while
still burning He in its core.
For a massive ($\approx1.0\ M_\odot$) and hot accreting WD, a shell
detonation can occur after the system has transferred only a few
hundredths of a solar mass of He \citep{Brooks15,Bauer17}, with the
donor orbital velocity $v_{\rm orb} \approx 600\ \rm km\ s^{-1}$.
In contrast, our $0.55\ M_\odot$ accreting WD model accumulates a
much larger ($0.23\ M_\odot$) He envelope before igniting when
${P_{\rm orb} = 9.4\ \rm min}$, while the donor has a much higher
orbital velocity of ${v_{\rm orb} = 880\ \rm km\ s^{-1}}$. These two
cases provide an estimate of the dynamic range in final velocities
from this scenario.

Depending on the orientation of ejection velocity $\vec{v}_{\rm f}$
relative to the system's prior orbit within the galaxy,
objects with $\vec{v}_{\rm f} \lesssim 800\ \rm km\ s^{-1}$ 
may either escape the galaxy or remain bound in unusual galactic
orbits (for initial galactic disk orbits of $250\ \rm km\ s^{-1}$ and
galactic escape velocity $550\ \rm km\ s^{-1}$, \citealt{Raddi19}).
Based on their trajectories from the galactic midplane, and under the
assumption that the objects they observed are young ($\leq 10\ \rm Myr$),
\cite{Raddi19} derived ejection velocities in the range $550\text{--}600\ \rm km\ s^{-1}$, with one object apparently orbiting counter to the rotation
of the galactic disk. For objects with trajectories that place them on
unbound orbits, they should leave the galaxy within
$(20\ {\rm kpc})/(500\ {\rm km\ s^{-1}}) \approx 40\ \rm Myr$ or
less. None of our models would fade and contract enough to reach the
WD cooling sequence within that amount of time (see
Figures~\ref{fig:radius} and~\ref{fig:HR}), and some even continue to
burn He in their cores. For objects that do remain bound to the
galaxy, they could eventually cool to form isolated WDs with masses
lower than any predicted from single-star evolution.

Due to tidal spin-up of the Roche-lobe filling donor stars, we also expect
that the runaway remnants should have significant rotational
velocities. Assuming tidal locking, we can estimate the initial
rotation period as $P_{\rm rot,0} = P_{\rm orb}$ using values from
Table~\ref{tab:donors}. There are then two possibilities that set
limits for the observable rotation velocities some time
later. In case~1, fluid elements conserve specific angular momentum,
so that rotation period at the surface will be
$P_{\rm rot} = P_{\rm rot, 0} (R/R_0)^2$,
where $R_0$ is evaluated in the initial model at the
mass coordinate corresponding to the final bound mass after stripping,
which is the fluid element that becomes the new stellar
surface. In case~2, we assume that the shear from such a rotation
profile would lead to angular momentum transport and rigid rotation,
so that total angular momentum conservation in the bound remnant sets
the rotation rate. The resulting period is
$P_{\rm rot} = P_{\rm rot,0} (I/I_0)$, where
$I = \frac{2}{3}\int_0^{M_{\rm f}} r^2 dm$ is the moment of inertia of
the mass that will remain bound after stripping.
Table~\ref{tab:rotation} gives surface rotation periods and velocities for
these two cases calculated using the \mesa\ models $10$ Myr after
explosion. Table~\ref{tab:rotation} also includes critical rotation fraction
$\Omega_{\rm rot}/\Omega_{\rm crit}$, where $\Omega_{\rm rot} = 2
\pi/P_{\rm rot}$ and $\Omega_{\rm crit} = \sqrt{GM/R^3}$.
Note that while we have used the profiles from our \mesa\ models to
make these estimates for rotational properties, we have not included
rotational effects in the \mesa\ models in this work.

We have not addressed the surface compositions of these remnant stars. Clearly, the dominant
species within the bulk of the donor star is the unburned helium, with
some carbon and oxygen from the earlier core burning. However, 
after the donor is stripped and still moving within the SN ejecta, it is 
expected that some of the low-velocity tail of the ejecta will 
be captured onto the remnant and heavily pollute its
surface \citep{Pan12b,Pan13}. 
Understanding the long-term mixing of these heavy elements in
He donor stars may help constrain timescales since explosion for
observable objects from this scenario. The opacities from
these elements may also influence the thermal evolution timescales as
long as they remain near the surface. These topics are left for future
work.

The methods applied here are also applicable to modeling 
binary systems where a WD is the donor star, which can lead to remnants
with even higher velocities \citep{Shen18}.
Equation~\eqref{eq:donor_radius} implies that the high-velocity
objects of \cite{Shen18} require significant expansion to reach their
currently observed radii, while Equation~\eqref{eq:pratios} suggests
this may be difficult to explain with shock heating of the donor
interior in the WD donor case. We are now exploring this case using 
\athena{} and \mesa\ models with the appropriate equation of state for 
degenerate WD interiors.

\acknowledgments
{\it Acknowledgments}:

We are grateful to the anonymous referee for a thorough report that
improved this paper. We thank Ken Shen, Dean Townsley, JJ Hermes,
Boris G{\"a}nsicke, and Tim Brandt for helpful discussions.
This research benefited from interactions with Jim Fuller, Abi Polin, and Eliot Quataert
that were funded by the Gordon and Betty Moore Foundation through Grant GBMF5076. 
We thank Bill Paxton for continuous efforts that enable the broad use of \MESA.
 This work was supported by the National 
 Science Foundation through grants
 PHY 17-148958 and ACI 16-63688.

\software{
  \mesa\ \citep{Paxton11,Paxton13,Paxton15,Paxton18,Paxton19},
  \athena,
  \texttt{Matplotlib} \citep{Matplotlib},
  \texttt{NumPy} \citep{Numpy}
}

\clearpage
\bibliographystyle{yahapj}

\end{document}